\input epsf
\magnification=1200  
\baselineskip 20pt
\def\prref{\par\noindent\hangindent=0.3cm\hangafter=1}

\def\pmb#1{\setbox0=\hbox{$#1$}%
  \kern-0.25em\copy0\kern-\wd0
  \kern.05em\copy0\kern-\wd0
  \kern-0.025em\raise.0433em\box0}

\def\prref{\par\noindent\hangindent=0.3cm\hangafter=1}

\def \cntl{\centerline}
\def \etal {{\it et al. }}
\def \cf {{\it cf. }}

\def \ie {{\it i.e. } }

\def \br {{ \bf r}}

\def \bk {{ \bf k}}

\def \bs {{ \bf s}}
\def \bd {{ \bf d}}
\def \bx {{ \bf x}}
\def \vd {{ \bf d}}
\def \bR {{ \bf R}}

\def \bepsilon {{\bf \epsilon}}
\def \WF {^{\rm WF}}
\def \CR {^{\rm CR}}

\def\rd{{\rm d}}

\def \apj {{\it Ap. J.}}

\def \ApJLet {{\it Ap. J. Lett.}}
\def \MNRAS  {{\it M.N.R.A.S.}}

\def \vjec {\vfill\eject}

\def \hmpc{h^{-1}Mpc}

\def \r0p{ r{_0^\prime}}
\def\prior{{\it prior}}

\def\today{\ifcase\month\or
  January\or February\or March\or April\or May\or June\or
  July\or August\or September\or October\or November\or December\fi
  \space\number\day, \number\year}
\nopagenumbers
\hfill \today 
\vskip 1.5 true cm
\cntl{\bf Non-linear Reconstruction of the Large Scale Structure}
\vskip 8pt
\cntl{\bf  V. Bistolas\footnote{$^1$}{E-mail: bisto@vms.huji.ac.il} and 
Y. Hoffman\footnote{$^2$}{E-mail: hoffman@vms.huji.ac.il}
}
\vskip 8pt
\cntl{ Racah Institute of Physics, The Hebrew University }
\cntl{Jerusalem, Israel}
\vskip 8pt
\cntl{\bf ABSTRACT}
\bigskip

The linear algorithm of the Wiener filter and constrained realizations (CRs) of Gaussian
random fields is extended here to perform non-linear CRs. The
procedure consists of: (1) Using linear CR of low resolution data to construct a  high
resolution  underlying field, as if the linear theory is valid; (2) Taking the linear
CR backwards in time, by the linear theory, to set initial conditions for an  N-body
simulation; (3) Forwarding the field  in time by an N-body code. An intermediate
step might be introduced to `linearize' the low resolution data.

The non-linear CR can be applied to any observational data set that is
linearly related to the underlying field.  Here it is applied to the IRAS 1.2Jy catalog
using 843 data  points within a sphere of $6000 Km/s$, to reconstruct the full
non-linear large scale structure of our `local' universe.

\vjec
\cntl{\bf I. Introduction}
\bigskip

\footline= {\hss\tenrm\folio\hss}
\pageno=2
In the standard model of cosmology galaxies and the large scale structure of
the universe form out of a  random perturbation field {\it via}
gravitational instability. It is assumed that  the primordial 
perturbation field constitutes a random homogenous and isotropic Gaussian field and that
on relevant scales its amplitude  is small, hence its evolution is described by
the linear theory of gravitational instability  (\cf\ Peebles 1980). The theoretical
study of structure formation has been a major effort of modern cosmology
(\cf\ Padmanabhan 1993). On the observational side, the large scale structure has been
studied mostly by means of red-shift surveys (\cf\  Strauss and Willick 1995) and
peculiar velocities (\cf\  Dekel 1994). A method
for the reconstruction of the underlying dynamical (density and velocity) fields from a
given  observational data base is presented here.

The problem of recovering the underlying field from   given observations,
which by their nature are incomplete and have a finite accuracy and resolution, is one
often encountered in many branches of physics and astronomy. It has been shown that
for a random Gaussian field  an optimal estimator of the underlying field is given by
a minimal variance solution (Zaroubi, Hoffman, Fisher and Lahav 1996; ZHFL), known also
as the Wiener filter (hereafter WF, Wiener 1949, Press \etal\ 1986). This approach is
based on the {\it a priori} knowledge of the statistical nature of the field, the
so-called \prior. Within the framework of Gaussian fields the WF coincides with the
Bayesian {\it  posterior}  and the maximum entropy estimations (ZHFL). Indeed, in the
cosmological case on large enough scales where linear theory applies and the (over)
density and velocity fields are Gaussian the WF is the optimal tool for the
reconstruction of the large scale structure. This is further complemented by the
algorithm of constrained realizations  (CRs) of Gaussian fields (Hoffman and Ribak 1991)
to create Monte Carlo simulations of the residual from this optimal estimation. This
combined WF/CR approach has been applied recently to a variety of cosmological data
bases in an effort to reconstruct the large scale structure. This includes the analysis
of the COBE/DMR data (Bunn \etal\ 1994), the analysis of the velocity potential
(Ganon and Hoffman 1993),
the reconstruction of the density field  (Hoffman 1993, 1994, Lahav
1993, 1994, Lahav \etal\ 1994) and the peculiar velocity field (Fisher \etal\  1995)
from the IRAS redshift survey (Fisher \etal\ 1993).

A major limitation of the WF/CR approach is that it applies only in the linear regime. 
Yet, on
small scales the perturbations are not small and the full non-linear gravitational
instability theory has to be used. Here  the WF/CR method  is extended to
the non-linear regime, and a new algorithm of non-linear constrained realizations
(NLCRs) is presented. The general method is presented in \S II and
its application to the IRAS 1.2Jy catalog is given in \S III. The results are presented
in \S IV and  a short discussion (\S V) concludes this {\it Letter}.

\bigskip
\cntl{\bf II. Non Linear Constrained Realizations}
\bigskip

The  general WF/CR method has been fully described in ZHFL and only a very short
outline of it is presented here.  Consider the case of a set of observations performed
on an underlying random field (with $N$ degrees of freedom) $\bs=\{s_1,...,s_N\}$
yielding $M$ data points, $\bd=\{d_1,...,d_M\}$. Here, only measurements that can be
modeled as linear convolution or mapping on the field are considered. The act of
observation is represented by  
$$
\bd = \bR \bs + \bepsilon,  \eqno(1)
$$
where $\bR$ is a linear operator   which represents the point spread function
and $ \bepsilon = \{\epsilon_i,...,\epsilon_M\}$ gives the statistical errors. Here the
notion of a point spread function is extended to include any linear operation that
relates the measurements to the underlying field. The WF  estimator is:
$$
\bs{\WF} = \Bigl<\bs \,\vd^\dagger\Bigr>\Bigl<\vd\,\vd^\dagger\Bigr>^{-1}
\vd \quad .\eqno(2)
$$
Here, $\Bigl<  ... \Bigr>$ represents an ensemble  average and  
$\Bigl<\bs \, \vd^\dagger\Bigr>$ is the cross-correlation matrix of the data and the
underlying field. The data auto correlation matrix is
$$ 
\Bigl<\bd \, \vd^\dagger\Bigr> =\bR \Bigl<\bs \, \bs^\dagger\Bigr> \bR + 
\Bigl<\bepsilon \, \bepsilon^\dagger\Bigr>,  \eqno(3)
$$
where the second term represents the statistical errors, \ie\    shot noise.
(See ZHFL for a general treatment of the error covariance matrix.)

In the case of a random Gaussian field, the WF estimator coincides with the conditional
mean field given the data. A  CR of the random residual from the mean is obtained by
creating an unconstrained realization of the underlying field ($ \tilde
\bs$) and the errors ($ \tilde \bepsilon$), and `observing' it the same way the actual
universe is observed. Namely, a mock data base is created by:
$$
\tilde \bd = \bR \tilde \bs + \tilde  \bepsilon,  \eqno(4)
$$
A CR is then obtained simply by (Hoffman and Ribak 1991):
$$
\bs\CR = \tilde \bs + \Bigl<\bs \,\vd^\dagger\Bigr>\Bigl<\vd\,\vd^\dagger\Bigr>^{-1}
\bigl( \bd   - \tilde \bd \bigr). \eqno(5)
$$

The WF/CR is used now to uncover the
finite   resolution  used to obtain the data. 
Thus, low resolution data is used to make high resolution  CRs. Here, for
simplicity the finite resolution is modeled by Gaussian smoothing, however any other
kernel can be used as well. The low resolution is modeled by   smoothing on scale $R_L$
and the high resolution by $R_S$, where $ R_L > R_S $. Often, the low resolution data
correspond to obseravables in, or close to, the linear regime, while the high
resolution field lies deep in the non-linear regime.
 However, the reconstruction is done within the
framework of the linear theory. This might be a reasonable assumption for the low
resolution data, but it is certainly inconsistent with the high resolution field.  The
following procedure is suggested here for the NLCR: I. Use low resolution data to
construct a high resolution CR, as if linear theory holds on these small scales; II.
Take this CR backwards    to an early enough epoch by the linear theory; III. Use this
CR, which now constitutes a given realization of the initial conditions of our `local'
universe constrained by   observational data, as an input to an  N-body code
to evolve it to the present time. These three steps provides one with a NLCR given the
observed data and the assumed \prior\  model. 

The three steps   proposed here are all consistent with the general
framework of the standard model, namely Gaussian primordial perturbation field and
 gravitational instability. Now, in the case where the
smoothing indeed transforms dynamical variables to the linear regime, the constructed
field provides one with a particular realization which is fully consistent with the
assumed model. The quality of the reconstruction  depends on the accuracy of the data
and its sampling and on the nature of the \prior\ , namely the `strength' of the
correlations. However, often it  happens that smoothing does not take the data all
the way to the linear regime. In such a case an intermediate step of  mapping  these
quasi-linear variables to the linear ones has to be introduced. Such a mapping cannot be
usually  rigorously formulated and one should recourse to some approximations. These
should be checked against N-body simulations and mock catalogs,   to find a mapping
suitable to the problem at hand.

\bigskip
\cntl{\bf III. Application: The IRAS 1.2Jy Catalog}
\bigskip

The WF/CR and the NLCR presented here can be used with any data base whose relation to
the underlying field can be modeled by Eq. 1. Thus, observations of the velocity field
can be used to reconstruct the density field and {\it vice versa}. The concrete case
studied here is the construction of the density and velocity fields from the IRAS 1.2Jy
redshift survey. At present red-shift distortions are ignored, however the formalism
can be easily extended to account for these as well (Zaroubi and Hoffman 1995). The
sample  is defined by its selection function, $\phi(r)$, and the boundaries of the 
survey are
defined by a mask of   galactic latitude   $\vert b \vert < 5^\circ$. The
\prior\ assumed here is  a CDM  power spectrum with a shape parameter $\Gamma=0.2$
and a normalization of $ \sigma_8 = 0.7$ (\cf\ Strauss and Willick 1995). For simplicity
no biasing and a flat universe are  assumed.

The underlying   density field is evaluated  on a Cartesian grid with a
sampling rate of $1000 Km/s$ (here distances are given in velocity units) within a
sphere of $6000 Km/s$, excluding IRAS' zone of avoidance. The discrete galaxy
distribution is   smoothed on a scale $R_L=1000 Km/s$. This yields $M=834  $
data points:
$$
\Delta_\alpha = \Delta(\br_\alpha) = \Bigl[ \sum_{\rm gal} 
 {1\over \phi(r_{\rm gal})   } 
 \exp\bigl( - { (\br_\alpha - \br_{\rm gal})^2 \over {2 R{^2_L}}  } \bigr) 
 - \bar n \Bigr] / \bar n   \eqno(6)
$$
where $ \bar n$ is the mean number density of the IRAS galaxies.
The data autocorrelation function is written as $ \Bigl< \Delta_\alpha
\Delta_\beta\Bigr> = \xi_{\alpha \beta} + \sigma_{\alpha \beta} $. The first term is
just the autocorrelation function of the smoothed field ( $\xi^s(r)$ ),
$$
\xi_{\alpha \beta} = \xi^s(\vert \br_\alpha - \br_\beta \vert ) 
= { 1 \over (2 \pi)^3 } \int P(k) \exp \bigl( -(k R_L)^2 \bigr) 
       \exp \bigl( {\rm i} \bk \cdot (\br_\alpha - \br_\beta ) \bigr) \rd^3k , \eqno(7)
$$
and the shot noise covariance matrix is: 
$$
\sigma_{\alpha \beta} = {1\over \bar n (2 \pi R_L^2)^{3/2} } 
\int { 1\over \phi(x) } 
\exp \bigl(- { (\br_\alpha-\bx )^2 + (\br_\beta-\bx )^2
\over
2 R{^2_L} }  \bigr)   \rd^3x     \eqno(8)
$$
Note that the   kernel introduces off-diagonal terms in the error covariance
matrix (ZHFL). The cross-correlation of the high resolution    field and the low
resolution data   is:  
$$
\xi_\alpha(\br_i) = { 1 \over (2 \pi)^3 } \int P(k) 
\exp \bigl( - { (k R_S)^2 + (k R_L)^2 \over 2 } \bigr) 
       \exp \bigl( {\rm i} \bk \cdot (\br_i - \br_\alpha ) \bigr) \rd^3k
\eqno(9)
$$
Defining the WF operator $W_{i\beta}$, 
$$  
W_{i\beta} =  \xi_\alpha(\br_i) \Bigr(\xi_{\alpha \beta} +
 \sigma_{\alpha \beta} \Bigr)^{-1},  \eqno(10)
$$ 
and a linear high resolution is thus obtained by
$$
\delta(\br_i) = \tilde \delta(\br_i) + W_{i\beta} 
\Bigl( \Delta_\beta - \tilde\Delta_\beta \Bigr).  \eqno(11)
$$

The choice of $R_L$ plays a crucial role  in the NLCR algorithm and it involves conflicting
considerations. On the one hand a small $R_L$ is desired, so as to keep high resolution
information, but on the other one  a large $R_L$ would minimize the shot noise errors
and would result in a `more' linear estimator. Here a value of $R_L=1000 Km/s$ is
chosen. To check the linearization of this   smoothing a   non-linear unconstrained
realization of the assumed \prior\ has been calculated by   a PM N-body code. Now, the
smoothed (scale $R_L$) (over)density is evaluated in two  ways. One is done by
smoothing the initial conditions and   propagating it in time by the linear theory,
yielding $\delta^L$. The other, $\delta^{NL}$,  is obtained   by smoothing the full
non-linear density field. It is
known that even on the $1000 Km/s$   smoothing scale there are systematic
deviation and scatter from the desired $ \delta^{NL} = \delta^L $ relation (Nusser
\etal\ 1991). One finds that on both ends of high amplitudes positive and negative
$\delta$'s,  $ \delta^{NL}$ is larger than   $ \delta^L $.
Note that this is a pure dynamical phenomenon and the statistical shot noise does not
affect it.   A minimal variance fitting formula is calculated here,   
$ \delta^L = f(\delta^{NL}) \delta^{NL}$, and this is used to recover the linear
field. Note that  $f(\delta^{NL})$ depends on the assumed model and the smoothing kernel.
A consistency check on this simple mapping is the evaluation of the 1-point 
distribution function of $f(\delta^{NL}) \delta^{NL}$. Indeed it is found to be
very close to that of $ \delta^L$, namely a normal distribution.

Various algorithms have been proposed to trace back non-linear
perturbation field to the linear regime (\cf\  Strauss and Willick 1995). All of these  `time
machines' recover the initial linear field in the case where the quasi-linear field is
known {\bf exactly}, with no statistical uncertainty. The case of real observational
data   where the shot noise 
 increases with distance, poses a much more difficult problem. As one goes
further away the data becomes more dominant by the shot noise and in the mean the
amplitude of the measured field increases with distance. Thus, before applying any `time
machine' the signal has to be first cleaned from the noise and only then it can be
traced back to the linear regime.

The  phenomenological fix to the `non-linearity' of the smoothed data 
which is used here consists of two steps. First, to account for the scatter in 
the ($ \delta^{NL}, \delta^L $) relation   a
new term is introduced to the data auto-covariance matrix, $\sigma^{NL}$. Dealing
with the scatter by statistical means is a manifestation of our inability to invert the
exact non-local non-linear mapping from the linear to the quasi-linear regime.  
Here we go to the extreme simplification and take 
$\sigma{^{NL}_{\alpha\beta}}=const. \delta_{\alpha\beta}$. The value of the constant
term is determine by the requirement that $\chi^2/d.o.f. = 1$, where the $\chi^2$
takes into account the cosmic variance, shot noise and  $\sigma^{NL}$. A WF estimator
of the $R_L$-smoothed field is obtained by applying a WF on the data, where $R_S$ is
replaced by $R_L$ to obtain low resolution, 
$$
\delta^{\rm WF,QL}(\br_i)=  \Biggl[ W_{i\alpha} \Biggr]_{R_S=R_L} \Delta_\alpha. 
\eqno(12)
$$ 
The estimation of the quasi-linear correction is given by 
$\bigl(1 - f(\delta^{\rm WF,QL}\bigr) \delta^{\rm WF,QL}$. This correction is evaluated
at grid points $\br_\alpha$ and is used to correct 
the data points:
$$
\Delta{^L_\alpha}=  \Delta_\alpha - 
\bigl(1 - f(\delta^{\rm WF,QL}\bigr) \delta^{\rm WF,QL}.
\eqno(13)
$$
The modified (`linearized') $\Delta{^L_\alpha}$'s are now substituted in Eq.  10 to
obtain a high resolution CR of the underlying linear field, given the actual data.

The linearization procedure presented here behaves as follows. In the limit of distant
data points, where the data is dominated by shot noise, the WF attenuates the estimated
field towards zero amplitude. Substituting the resulting estimator in Eq.  12 would
hardly change its value. The WF/CR is therefore dominated by the  random residual,
and consequently the resulting realization lies in the linear regime. For nearby data
points  where the shot noise is negligible,  the WF leaves the signal almost untouched
$ \delta^{\rm WF,QL} \approx \Delta $. In such a case the fitting formula would
linearized the data, as has been checked against the N-body simulations.

\bigskip
\cntl{\bf IV. Volume Limited IRAS Catalog}
\bigskip

The IRAS 1.2Jy catalog consists of 5321 galaxies These are used to evaluate the
smoothed density field on a Cartesian grid of $1000 Km/s$ spacing within a $6000
Km/s$, excluding the zone of avoidance, yielding 844 data points.  NLCRs are
created on a finer
$64^3$ grid of $250 Km/s$ spacing. A PM N-body code that is   used here  is based on
an FFT Poisson solver. For a CDM-like power spectrum the
structure within the $6000 Km/s$ would be hardly affected by the periodicity on the $\pm
8000 Km/s$ box. A comprehensive analysis of NLCR, including detailed comparison of
reconstruction of mock catalogs and sensitivity to the assumed prior has been conducted
and will be presented in a forthcoming paper (Bistolas and Hoffman, 1995a)

The IRAS galaxy distribution is presented in  Fig. 1, where the projected galaxy
distribution (within $\pm 1000 Km/s$) on the nine planes of $SGX, SGY, SGZ= \pm3000, 
0  Km/s$ is given. The full N-body distribution of the NLCR is
presented in Fig. 2 in a manner similar to that of Fig. 1. Note that this consists a
realization of a `volume limited' IRAS  catalog. Finally, the non-linear reconstructed
field   at $500 Km/s$   smoothing is presented in Fig. 3.  A full analysis of the
cosmography revealed by the NLCR will be given in a forthcoming publication (Bistolas
and Hoffman 1995b). Here we just point to the seemingly filamentry structure of the
reconstruced galaxy distribution. A closer inspection  shows that the generic feature
here is more   planar (2D)  rather than a filamentry (1D) structure, and the
apparent filaments are the intersection  of the sheets with the planes defined by the
plots.  Different NLCRs have been performed to study the
variance implied by  the \prior\  and the data and  relatively small scatter is
found between the different realizations.  In particular,
the existence and location of peaks and troughs is a very robust feature of the
realization with small scatter in their amplitudes. Also the sheets and filaments remain
invariant under the different realizations, however their `sharpness' varies somewhat.
 A comparison of two such realizations is
given in Fig. 4, where the `volume limited' galaxy distribution and $500 Km/s$  
smoothed $\delta$-field at the supergalactic ($SGZ=0$) plane are plotted.

\bigskip
\cntl{\bf V. Discussion}
\bigskip

The NLCR algorithm presented here enables one to perform controlled Monte Carlo
N-body simulations of  the formation of our `local'   universe. These are designed
  to recover the actual observational data, used to constrain them, within the
statistical uncertainties of the data. The new ingredient introduced here is the
reconstruction of the non-linear regime, \ie\ the extrapolation in Fourier space from
small to large wavenumbers that are deep in the non-linear regime.

The NLCR introduced here can serve as a tool for studying and analyzing the large scale
structure of the universe. Some of the obvious problems where NLCRs are expected to be
very useful are: (1) The reconstruction of the  velocity field from
redshift catalogs; (2) Mapping the zone of avoidance and extrapolating the dynamical
fields into unobserved regions; (3) Studying the dynamics   of
actually observed rich clusters with  the actual initial and boundary conditions; (4)
Analysis of filaments and pancakes as probes of   the initial conditions
and the cosmological model; (5) The NLCR can serve as a probe of the biasing mechanism.
The main virtue here lies in the fact that different data sets, which in principal can
represent different biasing of the underlying dynamical field, can be used to
simultaneously set constraints on the realizations.  Given all these and the technical
simplicity of the algorithm we expect it to be a standard tool of N-body and gas
dynamical simulations.

At the time this {\it Letter} has been written Kolatt \etal\ (1995)  have reported on  a
similar project of NLCR of the IRAS 1.2Jy catalog. Their procedure differs from the
present one mainly in not distinguishing between the low resolution (data) and high
resolution (realizations). The input data is smoothed on the $500 Km/s$ scale and is
heavily dominated by the noise, which is `removed' by a power preserving modified WF.
The modified filter is designed to preserve the power, regardless of the noise level.
The resulting estimator is therefore more dominated by the noise and less by the
\prior\ model compared to our method.   Yet, both methods seem to yield similar results
and are equally efficient. Detailed comparisons against N-body simulations are needed
to judge the merits of each method.

\vjec

\bigskip
\cntl{\bf   Acknowledgments}
\bigskip

The members of the IRAS collaboration are gratefully acknowledged for their help with the
IRAS data base. We have benefited from many stimulating discussions with L. da Costa, A.
Dekel, O. Lahav and S. Zaroubi. This research has been supported in part  by  The Hebrew
University Internal Funds (grant 53/94) and by the Israel Science Foundation
administrated by  the Israeli Academy of Sciences and Humanities (grant 590/94).

\vjec

\bigskip
\cntl{\bf References}
\bigskip

\prref Bistolas, V. and Hoffman, Y., 1995a (in preparation).

\prref Bistolas, V. and Hoffman, Y., 1995b (in preparation).

\prref Bunn, E., Fisher, K.B., Hoffman, Y., Lahav, O., Silk, J., \&
Zaroubi, S. 1994, \ApJLet , {\bf 432}, L75.

\prref Dekel, A., 1994, {\it Ann. Rev. Astron. Astrophys.}, {\bf 32}, 371.

\prref Fisher, K.B., Lahav, O., Hoffman, Y., Lynden-Bell, D. \& Zaroubi,
S. 1994,  \MNRAS, in press.

\prref Ganon, G. and Hoffman, Y., 1993, \ApJLet, {\bf 415}, L 5.

\prref Hoffman, Y. 1993, Proc. of the
${\rm 9^{\rm th}}$ IAP Conference on {\it
Cosmic Velocity Fields}, eds. F. Bouchet and M. Lachi\'eze-Rey,
(Gif-sur-Yvette Cedex: Editions Fronti\'eres), p. 357

\prref Hoffman, Y. 1994, in `{\it Unveiling
Large Scale Structures Behind the Milky-Way}', eds. C. Balkowski and R.C. Kraan-Korteweg, PASP conference series.

\prref Hoffman, Y. \& Ribak, E. 1991, \ApJLet, {\bf 380}, L5.

\prref Kolatt, T., Dekel, A., Ganon, G., and Willick, J.A., 1995, \apj (submit,).

\prref Lahav, O. 1993, Proc. of the ${\rm 9^{\rm th}}$ IAP Conference
on {\it Cosmic Velocity Fields}, eds. F. Bouchet and M.
Lachi\'eze-Rey,(Gif-sur-Yvette Cedex: Editions Fronti\'eres) p. 205

\prref Lahav, O. 1994, in `{\it Unveiling
Large Scale Structures Behind the Milky-Way}', eds. C. Balkowski and R.C. Kraan-Kortew
eg, PASP conference series.

\prref Lahav, O., Fisher, K.B., Hoffman, Y., Scharf, C.A., \& Zaroubi,
S. 1994, \ApJLet, {\bf 423}, L93.

\prref Nusser, A., Dekel, A., Bertschinger, E., and Blumethal, G.R.,   1991, \apj, {\bf
379}, 6.

\prref Padmanabhan, T., 1993,  Structure Formation in the Universe, (Cambridge:
Cambridge University Ppress).

\prref Peebles, P.J.E. 1980, The Large-Scale Structure of
the Universe, (Princeton: Princeton University Press).

\prref Press, W.H., Teukolsky, S.A., Vetterling, W.T., \& Flannery, B.P.
1992, Numerical Recipes (Second Edition) (Cambridge: Cambridge
University Press).

\prref Strauss, M.A. and Willick, J.A., 1995, {\it Physics Report} (in press).

\prref  Wiener, N. 1949, in {\it Extrapolation and Smoothing of Stationary
Time Series}, (New York: Wiley)

\prref Zaroubi, S., Hoffman, Y.,  Fisher, K.B., and S. Lahav, O., 1995, \apj,
(in press; ZHFL).

\prref Zaroubi, S., and  Hoffman, Y., 1995, \apj, (submit.).

\vjec

\bigskip
\cntl{\bf Figure Captions}
\bigskip

\prref Fig. 1: Raw data: The IRAS 1.2Jy galaxies. The galaxy distribution is presented
in 9 planar slabs of thickness of $\pm 10\hmpc$. (Supergalactic coordinates are used
and distances are given in $\hmpc$, where $h$ is Hubble's constant in units of $100
Km/s/Mpc$.)

\prref Fig. 2. `Volume limited' IRAS catalog: A non-linear constrained realization
based on the IRAS galaxy distribution. The full N-body particle distribution has been
diluted to the mean IRAS galaxies mean number density. 

\prref Fig. 3.  Gaussian smoothing: The non-linear constrained realization shown in
Fig. 2 is Gaussian smoothed on a scale $R=500 Km/s$. Contour spacing is $0.2$ and the
dashed lines correspond to negative values of $\delta$.

\prref Fig. 4. Different realizations: A comparison of   different non-linear 
constrained realizations of the same input data and \prior\  is presented here by the
`galaxy' distribution and the contour  plots. Upper and lower raws  correspond to two
different realizations. The galaxy distribution and contour plots are presented in the
same way as in Figs. 2 and 3.

\bye

\prref 

\vjec

\nopagenumbers

\epsfysize=550
\epsffile{FIG1.ps}

\epsfysize=550
\epsffile{FIG2.ps}

\epsfysize=550
\epsffile{FIG3.ps}

\epsfysize=550
\epsffile{FIG4.ps}

\end

 \bye
\nopagenumbers

\includegraphics{pic1b.ps}

\includegraphics{pic1d.ps}

\includegraphics{pic1c.ps}

\includegraphics{pic1a.ps}

\end